# Automation Will Set Occupational Mobility Free: Structural Changes in the Occupation Network

Soohyoung Lee[a], Dawoon Jeong[b,c*], Jeong-Dong Lee[a]

[a]*Technology Management, Economics, and Policy Program, Seoul National University, 1 Gwanak-ro, Gwanak-gu, Seoul 08826, Korea.*

[b]*Kellogg School of Management, Northwestern University, Evanston, IL, 60208, USA*

[c]*Northwestern Institute on Complex Systems, Evanston, IL, 60208, USA*

\* Corresponding author

E-mails: 2wkd0126@snu.ac.kr (Soohyoung Lee), dawoon.jung@kellogg.northwestern.edu (Dawoon Jeong), leejd@snu.ac.kr (Jeong-Dong Lee)

**Abstract**

Occupational mobility is an emergent strategy to cope with technological unemployment by facilitating efficient labor redeployment. However, previous studies analyzing networks show that the boundaries to smooth mobility are constrained by a fragmented structure in the occupation network. In this study, positing that this structure will significantly change due to automation, we propose the skill automation view, which asserts that automation substitutes for skills, not for occupations, and simulate a scenario of skill automation drawing on percolation theory. We sequentially remove skills from the occupation–skill bipartite network, and then investigate the structural changes in the projected occupation network. The results show that the accumulation of small changes (the emergence of bridges between occupations due to skill automation) triggers significant structural changes in occupation network. The structural changes accelerate as the components integrate into a new giant component. This result suggest that automation mitigates the bottlenecks to smooth occupational mobility.

## 1. Introduction

The anxiety surrounding technological unemployment—that labor will be substituted by machines and jobs will be lost—has persisted alongside continuous technological changes. Keynes [1] stated that technological unemployment occurs "due to our discovery of means of economizing the use of labor outrunning the pace at which we can find new uses for labor." This provides insight into potential solutions—technological unemployment can be alleviated by swiftly identifying new uses for labor and facilitating labor redeployment.

Through network analysis, previous studies have proposed occupational mobility as a strategy to cope with the threat of technological unemployment [2,3], a topic that has long interested innovation scholars, computational social scientists, and economists [4–17]. However, their findings are confined to showing that workers in highly automatable occupations are particularly vulnerable, as the occupations to which these workers can easily transition, based on their similar nature, are also prone to automation. Occupation networks in the literature often represent a fragmented structure based on the similarity between the skillsets of each occupation, thereby indirectly implying the existence of bottlenecks to smooth mobility across dissimilar occupations [3,18,19].

The approach of projecting the job market as a static occupation network has several limitations. Scholars agree that the similarity between the skillsets of occupations constrains occupational mobility. However, whereas the network can show the constraints of occupational mobility, the cause remains implicit [18]. An occupation is a bundle of skills [4,6–8,20–24] and skills are heterogeneously distributed across occupations [24–26]. Hence, the distribution of skills should be investigated alongside the structure of the occupation network as this would clearly reveal the bottlenecks to smooth mobility.

A growing body of literature has discussed the alteration of skillsets, or the bundle of skills, in occupations due to technological progress [4,7,22,24]. Researchers are interested as to how the accumulation of these changes, alongside ongoing automation, will eventually trigger significant changes in labor trends [4]. Nonetheless, occupation network studies have overlooked the potential for changes in the structure of the occupation network. One crucial reason for this limitation is that studies have relied directly on the automatability estimates from the seminal paper by Frey and Osborne [5], overlooking that these estimates are based on the likelihood of skill automation. To overcome this limitation, it is necessary to review the considerations of seminal papers [5,6] on what automation actually substitutes.

Access to skills is imperative to providing a first indication of what future labor dynamics will look like. By highlighting skills, this study aims to simulate the skill automation scenario by sequentially removing skills with high automatabilities from the occupation–skill bipartite network. Then, we can observe the structural changes in the projected occupation network. This methodology is based on the existing literature that investigates the mechanisms of structural change in networks. These include the phase transitions of interest in percolation theory [27–29], as nodes are sequentially removed [30,31]. The present study illustrates the mechanism by which automation of specific skills, the bottleneck to smooth mobility, triggers significant structural changes in the occupation network in the scenario. Through this approach, we expect to understand the impact of automation on labor trends by proposing the skill automation view, which asserts that automation substitutes skills rather than occupations. Whereas previous scholars have shown interest in labor trends transformation due to skill automation [4], to the best of our knowledge, no other study has thoroughly analyzed the mechanisms of the structural change in the occupation network.

This study contributes to the literature in two ways. Theoretically, our findings extend the discussion of how micro-level changes in skillsets can lead to macro-level changes in labor trends [4]. We also contribute to the convergence of network and labor studies by applying the methodology of network simulation to occupation network based on O*NET data from the United States. The skill automation scenario provides a framework to comprehend the mechanisms of how automation will change the labor trends. Practically, this study suggests a new tool for individual workers to gauge how the job market will change. By understanding that automation substitutes for skills rather than occupations, new pathways can be revealed in the process of structural changes in occupation network. Additionally, we extend the results of this study into discussions that derive labor policy implications to cope with the technological changes.

The remainder of this paper is organized as follows. Section 2 reviews previous studies and proposes the skill automation view, Section 3 describes how the occupation network is constructed using O*NET data and delineates the skill automation scenario process, Section 4 presents the findings of the study, and Section 5 discusses the implication of this study. Section 6 concludes the paper.

## 2. Literature Review

### *2.1 Previous studies on occupation network and specific skills*

The relationship between the skillsets of various jobs, often termed skill relatedness [32], has two facets with respect to job opportunities. Whereas workers can find the opportunities for career development, wage increases, and tenure, utilizing skill relatedness [23,33,34], the boundaries of job transitions are constrained by the relatedness [18,35], making it difficult to find new uses for labor due to inertia [36].

Previous studies suggest occupational mobility as a strategy for workers to cope with technological unemployment. They attempt to find alternative opportunities for vulnerable workers by investigating the network that represents the relatedness between occupations. Simultaneously, they highlight workers who may encounter greater difficulties due to the structure of occupation networks, which are fragmented into multiple components. For example, Dworkin [2] assessed the automatabilities of each occupation by the components in occupation network.[1] He suggested that industrial workers mostly face challenges in escaping the threat of automation, because the alternative occupations available to them are also prone to automation. Christenko [3] investigated each component in the occupation network and argued that workers lacking adequate skills for safe occupations cannot transition to them.

While examining the structure of an occupation network is necessary, an approach that focuses on occupations may be limited [4,22]. It is necessary not only to observe the relatedness between occupations but also to emphasize the heterogeneity of the skill bundles that constitute each occupation [4,25,26]. An occupation comprises a bundle of tasks [6–8,20–23], and a skill represents the capability required by a worker to perform tasks [8,20,22,24]. In this context, an occupation should be considered as the bundle of skills. Therefore, not only occupations but also skillsets should be simultaneously highlighted to thoroughly understand labor dynamics such as occupational mobility [4,35].

Human capital scholars discuss the role of specific and general skills in job transitions. Lazear [25] stated that all skills are common, but the weight of their importance is heterogeneous across all occupations [26] that is variously distributed across occupations. A skill with limited application [33], which is productive only within similar occupations [34], can be considered a specific skill. Thus, the specific skills are exclusively shared by similar occupations at comparable levels. In other words, a specific skill can be defined as a skill that is required for a relatively limited number of occupations [37], leading to a lock-in of mobility [33]. Given that occupational mobility is determined by skill relatedness [2,3,35], workers can smoothly transition between similar occupations utilizing specific skills, but may not transition across dissimilar occupations, that is, specific skills constrain occupational mobility [19]. Therefore, to deeply understand the structure of occupation network, one needs to identify how specific skills are distributed across similar or dissimilar occupations.

---

[1] Business, scientific, medical, service and industrial components

To analyze occupational mobility through a network-driven approach, Dworkin [2] and Christenko [3] projected the occupation–skill network onto the occupation mode, whereas Alabdulkareem et al. [35] and Hosseinioun et al. [37] projected it onto the skill mode. Expanding on these studies, we argue that directly observing the occupation–skill network would provide a clearer understanding of the function of specific skills in occupational mobility, based on the discussion in human capital literature. Therefore, this study first examines the projected occupation network and then investigates the occupation–skill network to clearly reveal how specific skills constrain the occupational mobility.

*2.2 Dynamics in Occupation Network and Target of Automation*

Frank et al. [4] described how automation impacts labor trends: "Technology typically performs specific tasks and may then alter demand for specific workplace skills. These micro-scale changes to skill demand can accumulate into systemic labor trends […]." A growing body of empirical studies show "micro-scale changes to skill demand." They suggest that the introduction of machines, such as robots [38,39] or artificial intelligence [40], leads to an increase in employment. These findings, which contrast with the conventional wisdom that the introduction of machines leads to a decrease in employment, can be explained by deskilling [5] or partial automation [41] of the bundle of skills. When the skill requirements of an occupation are simplified due to automation [21], transition into that occupation becomes easier.

However, few models have illustrated the mechanisms by which these changes "accumulate into systemic labor trends" [4]. Although network analysis can be a useful tool for explaining these mechanisms, existing occupation network studies are limited in modeling the labor trends induced by alterations in specific skills. Synthesizing percolation theory and fundamental considerations in technological unemployment literature can provide useful insights to overcome the limitations in existing studies.

Percolation theory examines the mechanism of the phase transition that occurs as small changes accumulate, eventually forming a larger trend [27–29], which can be referred as structural changes in a network. Chang et al. [27] illustrated the mechanism of network structural changes by simulating the emergence of a giant component through the integration of small components with a small but growing number of bridges, which aligns with percolation theory. They highlighted that information diffusion accelerates when exchanges of inventors take place across research teams within a company through managerial decisions. This can be considered as integration of fragmented components in a network due to external shock. They elucidated this phase transition in the network as a phenomenon in which small changes accumulate, eventually accelerating significant structural changes.

To model the simulation of structural changes in the occupation network caused by automation, reviewing the seminal literature on what automation substitutes for is crucial. Drawing on the Cobb-Douglas production function, Autor et al. [6] elucidated that the demand for routine "skills" decreases as the cost of computing power, which has a comparative advantage in routine tasks, decreases. In their widely cited research, Frey and Osborne [5] developed Autor et al.'s [6] model by utilizing nine O*NET variables, which represent skills, to estimate the automatabilities of occupations. These studies fundamentally considered the impact of machines on skills rather than occupations. Labor is the input for production, and workers utilize their skills, which are capabilities to perform tasks [4,8,20,22,24], to produce goods. Thus, labor in the models of Frey and Osborne [5] and Autor et al. [6] should be understood as skills rather than occupations. In other words, automation substitutes for skills, not occupations. In line with this, the automatabilities of occupations estimated by Frey and Osborne [5] need to be translated into the automatabilities of skills.

Following Frank et al. [4], alteration of specific skills due to automation can be postulated as an external shock to the occupation network. The findings from Alabdulkareem et al. [35] reveal that skills with strong relationships to others, such as basic skills, social skills, and cognitive abilities, have a greater influence on occupational mobility than certain types of skills, such as physical abilities, psychomotor abilities, and knowledge. The latter were monumentally identified as specific skills by Hosseinioun et al. [37], who illustrated the dependency of skills through the hierarchical structure of the skill network. Considering that Frey and Osborne [5] suggested that the engineering difficulty of perception and manipulation would be overcome with relatively fewer challenges, physical and psychomotor abilities, which are a subset of specific skills, are prone to automation.

We propose the skill automation view, which defines labor as skills, not occupations. Drawing on Frank et al. [4], we argue that the structure of occupation network will significantly change due to skill automation. Before automation, the occupation network has a fragmented structure because the specific skills exclusively required by similar occupations define the boundaries of mobility. If automation makes such skills no longer necessary, the bundles of skills in each occupation will change. The accumulation of these small changes leads to the occupation network evolving into a sticky structure as the bottlenecks to occupational mobility disappear.

## 3. Material and Methods

### *3.1 Occupation Network*

Extant studies represent the network showing relatedness among activities in various dimensions for analysis [42–45]. The methods of these studies are based on Hausmann and Klinger's [46] idea that if two activities co-occur frequently, they require similar capabilities. The occupation network literature has adopted this approach and the co-occurrence has been translated into transition potential between occupations [3,8]; the more overlapped the required skills of two occupations, the higher the transition potential.

By employing this approach, we construct the occupation network by utilizing the data that examine the importance of 120 skills for each occupation in the O*NET database.[2,3] The nodes in the network represent occupations, and the links represent the transition potential between occupations via overlapping skills, that is, occupational mobility. Studies have illustrated occupational mobility symmetrically based on similarity [2,3,8,17,23]. Although the occupation similarities can be symmetrical, the transition potential is inherently asymmetric. This asymmetry arises from a shortage or redundancy in skill sets relative to the target occupation, often termed skill mismatch [16,20]. Moreover, an asymmetric measure of relatedness can offer new insights [47]. Therefore, modifying Hidalgo et al. [42] and Alabdulkareem et al. [35], we utilize conditional probability to represent the asymmetric transition potential $T_{o,o\prime}$, as follows:

$$T_{o,o\prime} = P[IM_{o,s} \geq 2.5 \mid IM_{o\prime,s} \geq 2.5] \qquad (1)$$

---

[2] See: https://www.onetcenter.org/db_releases.html (O*NET 25.1)
[3] We consider knowledge, skills, and abilities (KSA) variables in O*NET as the skill following Nedelkoska et al. [8] and Hosseinioun et al. [37].

where $T_{o,o\prime}$ is the conditional probability and represents the overlap source occupation $o$'s skill requirements given target occupation $o\prime$'s. $IM_{o,s}$ is the importance of each skill for occupation $i$, ranges from 1 (not important at all) to 5 (very important). Based on Nedelkoska et al. [8], we postulate that a skill is required to an occupation if its importance for an occupation is greater than or equal to 2.5. Therefore, $T_{o,o\prime}$ reflects the ratio of skill requirements for both occupations to skill requirements for target occupation. For example, if occupation $o\prime$'s skill requirements are $s_1, s_2, s_3, s_4, and\ s_5$ and occupation $o$'s skill requirements are $s_3, s_4, and\ s_5$, $T_{o,o\prime} = 0.6$.

### *3.2 Skill Automation Scenario*

Albert, Jeong, and Barabasi [30] investigated the structural changes in a scale-free network that occurs when nodes are removed in the order of the highest degree. Using this method, Jo et al. [31] examined the COVID-19 infection network and observe that nodes representing infected individuals disappear rapidly from the network when high-degree nodes are removed. The method, which explores changes in unipartite networks, is employed in ecology studies observing changes in bipartite networks. The ecology literature examines the extinction of species, which are the nodes on one axis of the network, as pollinators [48], or habitats [49] (which are the nodes on the other axis of the network), disappear.

Extending these approaches, we implement the skill automation scenario to observe structural changes in the projected occupation network as the removal of skills on the occupation–skill bipartite network. Fig. 1 is a schematic representation of the skill automation scenario. The upper panel indicates the removal of skills on the occupation–skill network, depicted as a matrix. Studies have investigated the occupation networks by directly applying Frey and Osborne's [5] automatabilities of occupations to the occupation nodes. However, following the skill automation view, we assign automatabilities to the skills, which is the row in the matrix. Sequentially removing the skill rows in the matrix, we project the occupation network using the updated $T_{o,o'}$ values (lower panel in Fig. 1) and repeat the process to observe structural changes at each time step.

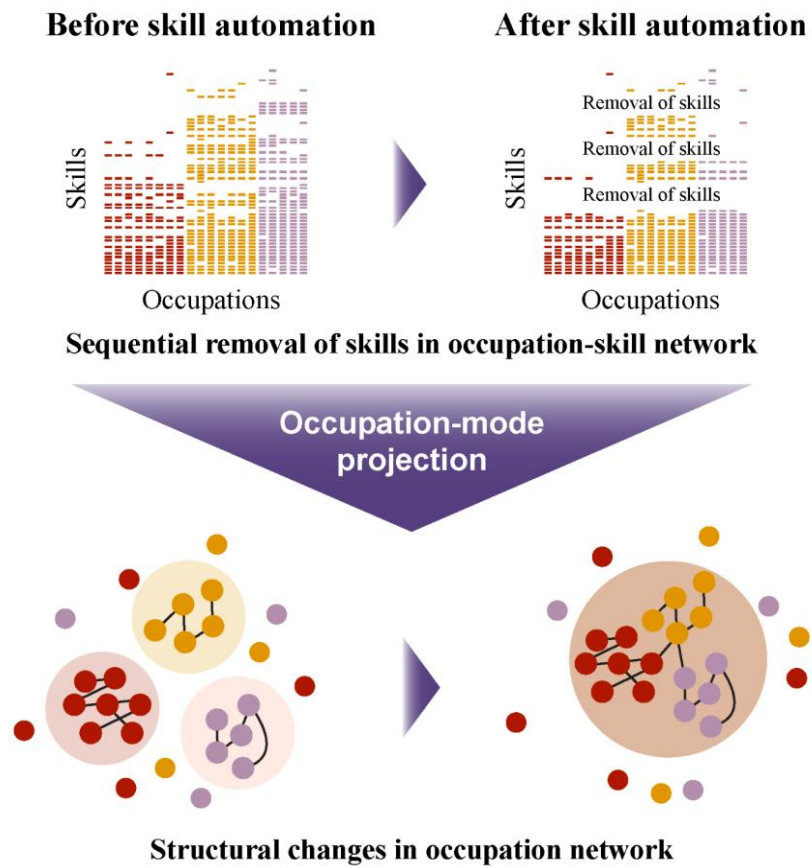

Fig. 1. The schematic representation of the skill automation scenario.

Frey and Osborne [5] estimated the probability of occupations being substituted by a comprehensive range of machines, including robots and artificial intelligence. Considering that skills important to highly automatable occupations are more likely to be automated, skill automatability can be determined using the correlation coefficient between automatabilities of occupations and the skill importance within each occupation [2,50].

$$automatability_s = corr(FO_o, IM_{o,s}) \qquad (2)$$

The sequence of skill removal is determined in descending order of $automatability_s$. Additionally, we assume that skills with higher importance ($IM_{o,s}$) are prioritized for removal, given their greater impact on occupational requirements. We implement the scenario until the skills with the positive $automatability_s$ values are eliminated.

We assess the relationship between automatability and specificity of each skill to explore which skills are substituted by automation. Alabdulkareem et al. [35] and Hosseinioun et al. [37] observed the contributions of each skill to labor dynamics through the analysis of skill networks. Simplifying their measure, we assume that specific skills are required by a relatively small number of occupations, and specificity can be represented as

$$specificity_s = -1 * \sum_o R_{o,s}$$
$$R_{o,s} = 1 \; if \; IM_{o,s} \geq 2.5, otherwise \; 0 \qquad (3)$$

$R_{o,s} = 1$ indicates that skill $s$ is required for occupation $o$. We standardize the $specificity_s$ as z-score and define the skill that has a value exceeding 0 as a specific skill.

*3.3 Dynamics in the occupation network: Strongly Connected Component (SCC)*

We examine structural changes in the occupation network focusing on the core–periphery structure through the skill automation scenario. SCCs are the core of a directed network; despite receiving less attention, they can serve as a valuable tool for assessing the structure in a directed network [51,52]. Several methods to delineate the core of an undirected network exist, but the occupation network in this study indicates asymmetric transition potentials between occupations. This requires an analysis that incorporates the directionality of links.

All nodes participating in an SCC form the stickiest connections within the network. They are mutually reachable regardless of the path taken. Paths originating from any node within an SCC possess cyclic paths that traverse all other nodes before returning to the starting node [51]. Therefore, the paths passing through the nodes in an SCC traverse without any bottlenecks. Nodes A, B, C, and D in the left panel of Fig. 2 illustrate the cyclic path in an SCC. Node D solely has an in-directional link from C and an out-directional link to B, whereas nodes A, B, and C share immediate two-way paths with each other. Nonetheless, node D participates in SCC 1 due to the cyclic path D-B-A-C-D. Consequently, nodes A and D participate in the same SCC, even without an immediate link. In summary, the cyclic path is necessary to establish an SCC, regardless of whether the links between participating nodes are immediate or mediated.

Nodes E and F show the importance of the cyclic path in forming an SCC. Node E may appear to be an influential node because it has two out-directional links to SCC 1. However, given that the paths starting from node E do not return to their origin, it can be considered structurally unrelated to SCC 1. Similarly, Node F may seem to be influenced by SCC 1 because it has two in-directional links from SCC 1, but for the same reason, it is structurally unrelated to SCC 1. Instead, node F participates in SCC 2, which consists of nodes F, G, and H.

Chang et al. [27] highlighted the role of bridges in integrating small components into a giant one and triggering significant structural changes in the network. The right panel of Fig. 2 exemplifies the structural change in a directed network where a single link integrates two old SCCs and triggers the emergence of a new giant SCC. The evolution of the one-way link, which represents the bottleneck of path between two SCCs, to a two-way link between nodes C and F generates the cyclic path F-G-H-F-C-D-B-A-C-F. Consequently, the fragmented structure of the network changes to a sticky structure without any bottlenecks in the new SCC.

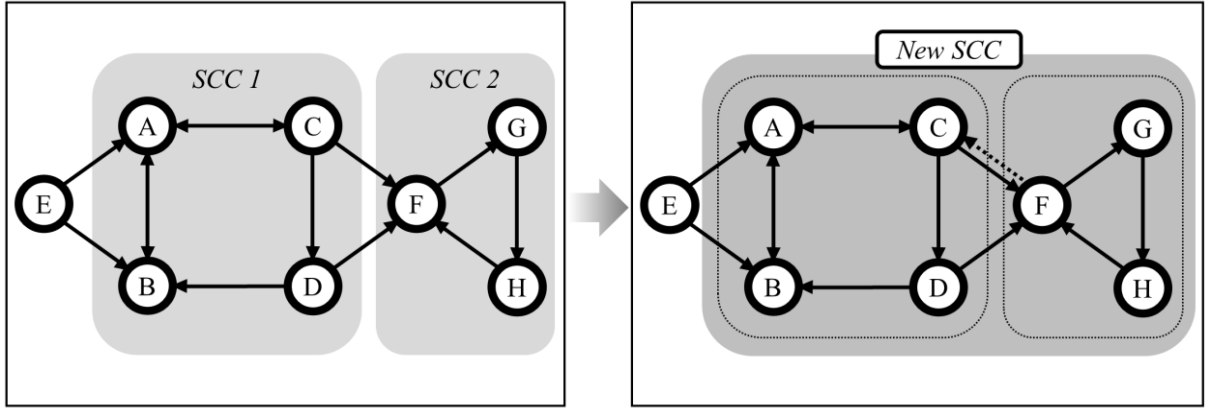

Fig. 2. The schematic representation of the cyclic paths in SCCs and the integration of two SCCs into a giant one.

An SCC in the occupation network represents the boundaries where the mobility is smooth. Assuming that a one-way link between two nodes in the network reflects a shortage or redundancy between the skillsets of the two occupations, the mobility that contradicts the directionality of the one-way link has a bottleneck. However, the sticky relationships between nodes participating in an SCC, where all paths are cyclic, indicate that there are no such bottlenecks on occupational mobility. Considering this, we explore the structural changes in the occupation network focusing on the SCCs.

## 4. Results

### *4.1 Determinants of Fragmented Structures in Occupation Network*

We construct the occupation network, comprising the directed links between occupations when $T_{o,o\prime}$ is greater than a certain threshold ($\theta = 0.94$), reflecting a real-world occupational mobility (Appendix A). Then, we set all weight of links to the same as 1 for simplicity. Equation 4 is mathematical representation of the directed and unweighted occupation network in Fig. 3.[4] $N_o$ is a set of 505 occupations and $L_{o,o'}$ is a set of directed links from occupation $o$ to occupation $o'$.

$$Occupation\ Network = G(N_o, L_{o,o'})$$

$$N_o = \{o_1, o_2, o_3, \ldots, o_{505}\} \tag{4}$$

$$L_{o,o'} = \{l_{o,o'} | l_{o,o'} = 1\ if\ T_{o,o'} \geq \theta, otherwise\ l_{o,o'} = 0\}$$

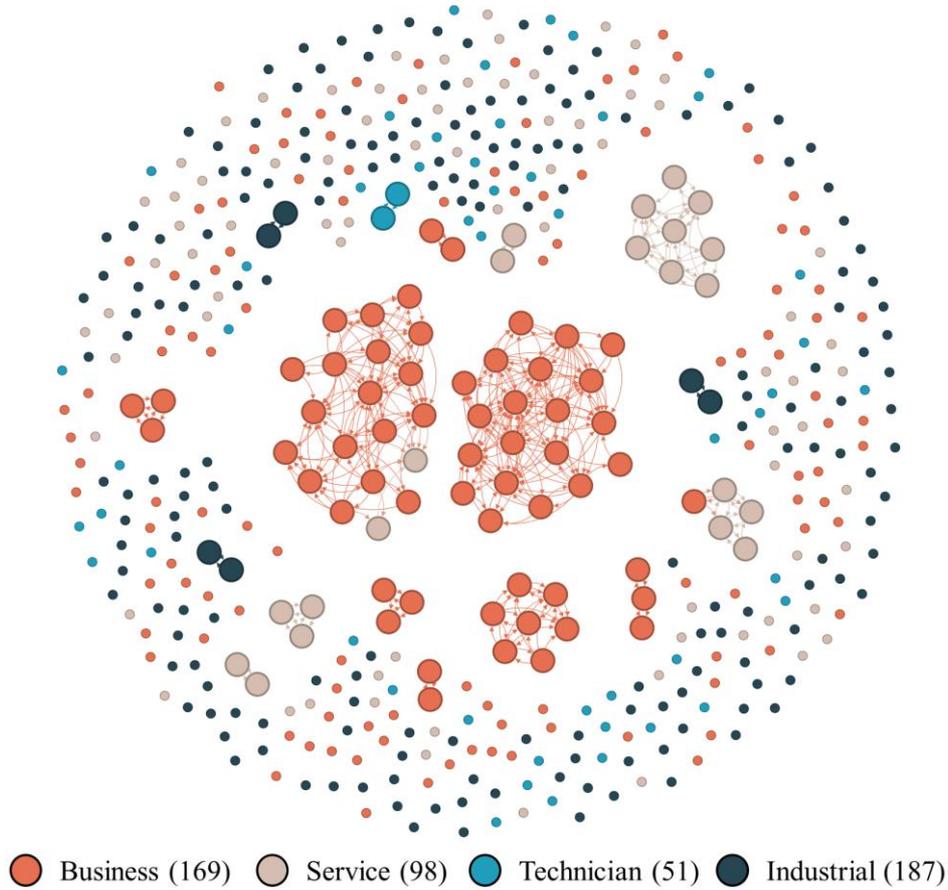

Fig. 3. An occupation network. The network is visualized using only the links that form the SCCs to highlight the structure.

---

[4] To coarsely categorize occupations based on their characteristics, we run the Louvain algorithm, an efficient algorithm for detecting optimal communities [2,35,51,53]. Drawing upon Dworkin [2], each community is named after the occupations that have the least deviation from the average skill importance value within that community. This approach facilitates more intuitive labeling based on the skill requirements of each occupation.

The network shows a fragmented structure consisting of several large and small SCCs, along with isolated occupations that do not participate in any SCC. Each SCC is composed of similar occupations. Table 1 provides an overview of occupations in the five largest SCCs. The giant SCC (GSCC) generally includes managers, whereas the second largest SCC includes financial occupations. The third, fourth, and fifth largest SCCs are formed by social care service, engineering, and medical occupations, respectively. The list of occupations within each SCC demonstrates that SCCs represent the boundaries of smooth mobility. SCCs predominantly consist of occupations with comparable skill requirements. For example, individual managers, who participate in the GSCC, may transition across various fields as they advance in their careers. However, a comparison of occupations across different SCCs suggests potential bottlenecks to occupational mobility between components. For instance, it is difficult for financial workers to transition into medical occupations, and vice versa. Therefore, the fragmented structure in occupation network reflects the potential bottlenecks to smooth mobility across dissimilar occupations.

**Table 1.**

The list of occupations that form the Top Five SCCs

| Communities | 2018 Census occupation title |
|---|---|
| GSCC (10 of 20 occupations) | Chief Executives; Advertising and Promotions Managers; Marketing Managers; Sales Managers; Human Resources Managers; Training and Development Managers; Purchasing Managers; Transportation, Storage, and Distribution Managers; Education and Childcare Administrators; Entertainment and Recreation Managers. |
| 2nd largest SCC (10 of 20 occupations) | Financial Managers; Compensation, Benefits, and Job Analysis Specialists; Fundraisers; Accountants and Auditors; Budget Analysts; Insurance Underwriters; Financial Examiners; Credit Counselors and Loan Officers; Tax Examiners and Collectors, and Revenue Agents; Tax Preparers. |
| 3rd largest SCC (8 occupations) | Clinical and Counseling Psychologists; Marriage and Family Therapists; Mental Health Counselors; Child, Family, and School Social Workers; Healthcare Social Workers; Social and Human Service Assistants; Preschool and Kindergarten Teachers; Special Education Teachers. |
| 4th largest SCC (7 occupations) | Aerospace Engineers; Chemical Engineers; Industrial Engineers, including Health and Safety; Mechanical Engineers; Nuclear Engineers; Engineers, All Other; Occupational Health and Safety Specialists and Technicians. |
| 5th largest SCC (5 occupations) | Optometrists; Surgeons; Podiatrists; Veterinarians; Nurse Practitioners. |

Note: The two largest SCCs have identical sizes, yet they are designated as the GSCC and the second largest SCC based on the progression of the subsequent scenario.

However, the particular cause that fragments the structure in the network remains unclear. Prior literature [18,19,26] has found that specific skills, which are required by a limited number of similar occupations [37], determine the boundaries of occupational mobility. In this respect, a more distinct cause for the fragmented structure in occupation network can be identified through the occupation–skill network in Fig. 4, which represents the skillsets of each occupation in the five largest SCCs as a matrix form. Equation 5 is the mathematical representation of the occupation–skill network in Fig. 4. $N_o$ is a set of 505 occupations and $N_s$ is a set of 120 skills. $L_{o,s}$ is a set of links between occupations and skills.

$$Occupation - Skill\ Network = G(N_o, N_s, L_{o,s})$$
$$N_o = \{o_1, o_2, o_3, \dots, o_{505}\}$$
$$N_s = \{s_1, s_2, s_3, \dots, s_{120}\}$$
$$L_{o,s} = \{l_{o,s} | l_{o,s} = 1\ if\ R_{o,s} = 1, otherwise\ l_{o,s} = 0\}$$

(5)

Fig. 4 shows that the specific skills exclusively shared within the five largest SCCs are heterogeneous, whereas certain skills are common across all occupations. "Medicine and Dentistry," "Biology," "Manual Dexterity," "Arm-Hand Steadiness," and "Control Precision" are skill requirements for medical occupations in the fifth largest SCC. A few occupations in other SCCs require those skills. Similarly, "Engineering and Technology," "Mechanical,"

"Operation Monitoring," "Quality Control Analysis," and "Design" are exclusively shared by engineering occupations in the fourth largest SCC, whereas these skills are seldom found in the skillsets of occupations in other SCCs. These heterogeneously distributed specific skills are shared within, but not between, occupations in each SCC. Thus, the specific skills can facilitate smooth mobility within each SCC, but can be bottlenecks to mobility across SCCs.

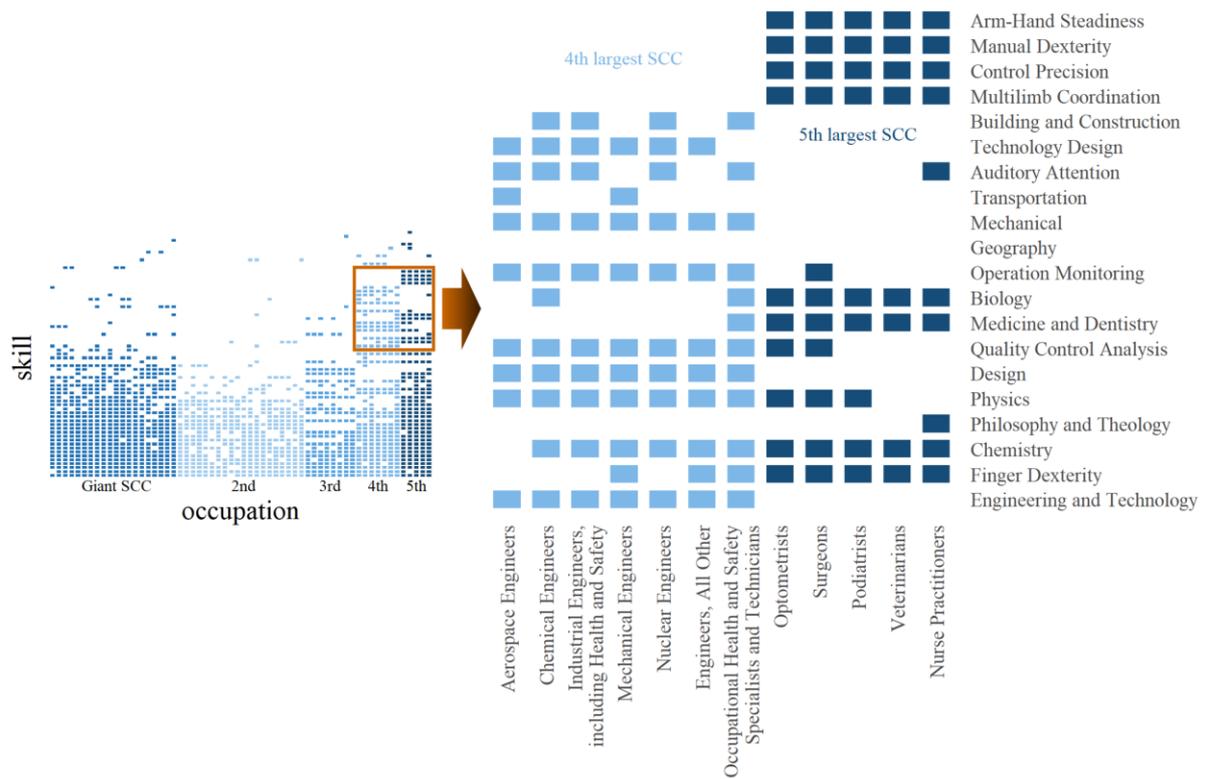

Fig. 4. The sample of the occupation–skill network illustrating the occupations in the five largest SCCs and subset of their skillsets. A tile indicates the link between a skill and an occupation ($R_{o,s} = 1$). The full network is represented in Appendix B.

In brief, by examining the occupation network, we identify similar occupations construct the boundaries of smooth mobility. Simultaneously, the specific skills exclusively shared by a small number of occupations, are heterogeneously distributed within the occupation–skill network. This indicates that the specific skills exclusively required to similar occupations, function as bottlenecks to mobility across SCCs, thereby determining the fragmented structure in an occupation network.

*4.2 Structural Changes in Occupation Network due to Skill Automation*

The skill automation scenario in this study aims to simulate the structural changes in occupation network by the sequential removal of skills and to explain the mechanism of these changes, by drawing on percolation theory [27–29]. Chang et al. [27] investigated the structural changes, which is the phase transition in the network, highlighting the emergence of a giant cluster. Drawing on this literature, we present the results of the skill automation scenario focusing on the GSCC growth.

Inset A in Fig. 5 shows that when the scenario progresses to 61.3%, the structure in the network insignificantly changes. The size of the GSCC, which represents the prominent smooth occupational mobility, changes very little. Business and service occupations primarily form relatively large SCCs, whereas technician and industrial occupations form small SCCs. Inset B in Fig. 5 illustrates the sudden growth of the GSCC when the scenario progresses to 64.8%. Furthermore, GSCC growth more significantly accelerates when the scenario progresses to 74.7% (Inset C). Two additional sudden growths follow throughout the scenario (Inset D and E).

The patterns in structural changes are different between the earlier stages leading up to the first sudden GSCC growth and that in the later stages. In Inset A, seven large SCCs with more than five nodes can be identified. These large SCCs survive the first sudden structural change (Inset B). By contrast, during the second sudden structural change, the number of those large SCCs decreases from seven to three (Inset C). At the same time, the growth of GSCC accelerates compared to the previous stage. Simultaneously, occupations from different categories start to merge into the GSCC.

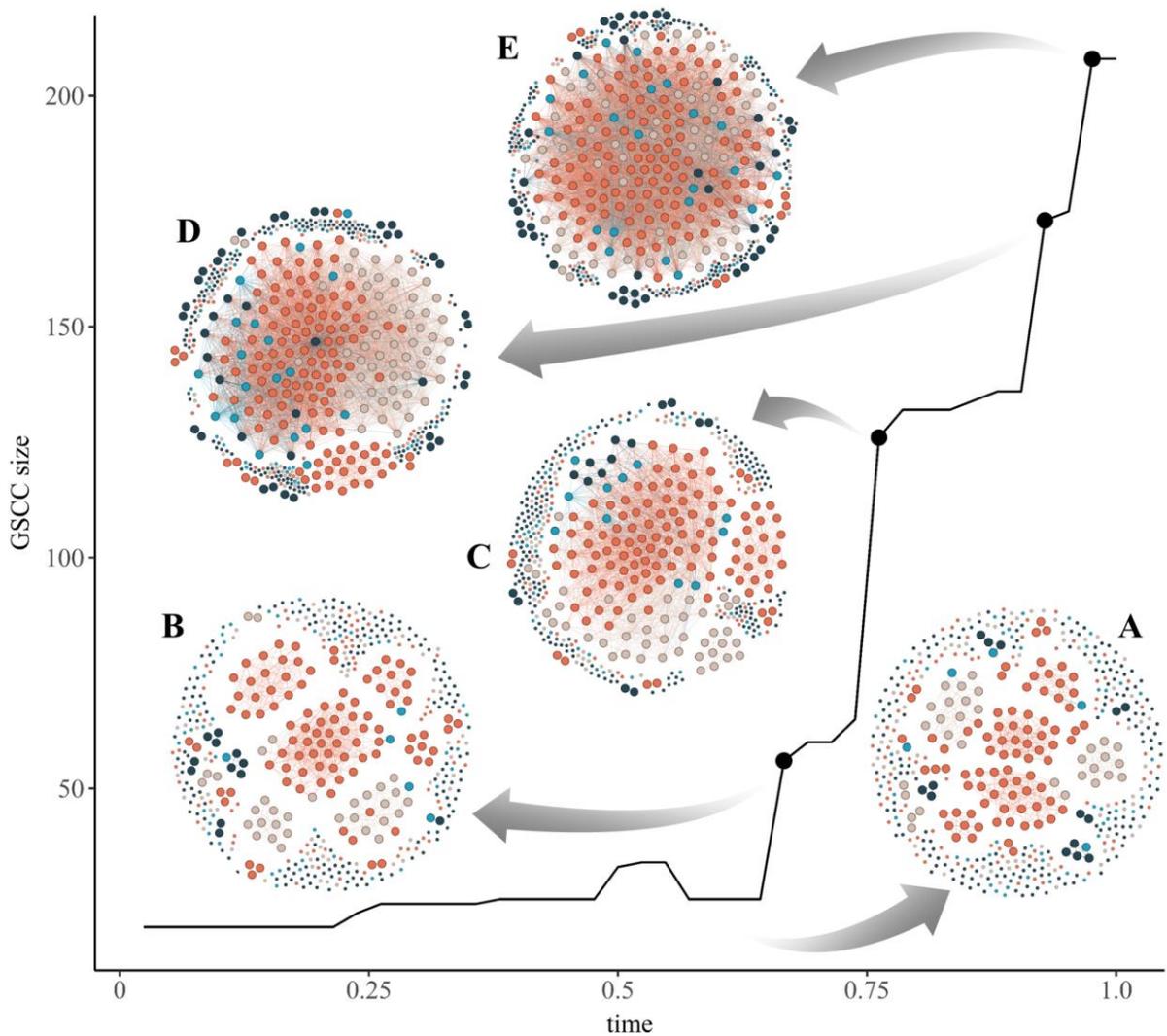

Fig. 5. Growth pattern of the GSCC and structural changes in occupation network. Closed dots on the line indicate the moments when the GSCC growth accelerates. See Fig. 3 for the categories of occupations in the networks. The networks are visualized using only the links that form the SCCs to highlight the structure.

This pattern is aligned with Chang et al. [27], who noted that adjacent nodes initially form multiple small components and eventually integrate with the largest one, thereby accelerating the largest component's growth. Fig. 6 confirms that the acceleration of the structural changes was caused by the integration of SCCs through the skill automation scenario.

The average size of small SCCs generally increases and fluctuates, whereas the number of SCCs, including isolated occupations, steadily decreases throughout the scenario. The general increase in the average size of small SCCs and steady decrease in the number of SCCs, indicate that small SCCs grow by integrating with isolated occupations. The notable moments are when the growth of small SCCs sharply declines. These coincide with the moments of sudden growth in the GSCC. Thus, the fluctuations in the average size of small SCCs indicate that multiple small SCCs grow and then integrate with the GSCC, contributing to the acceleration of the structural changes.

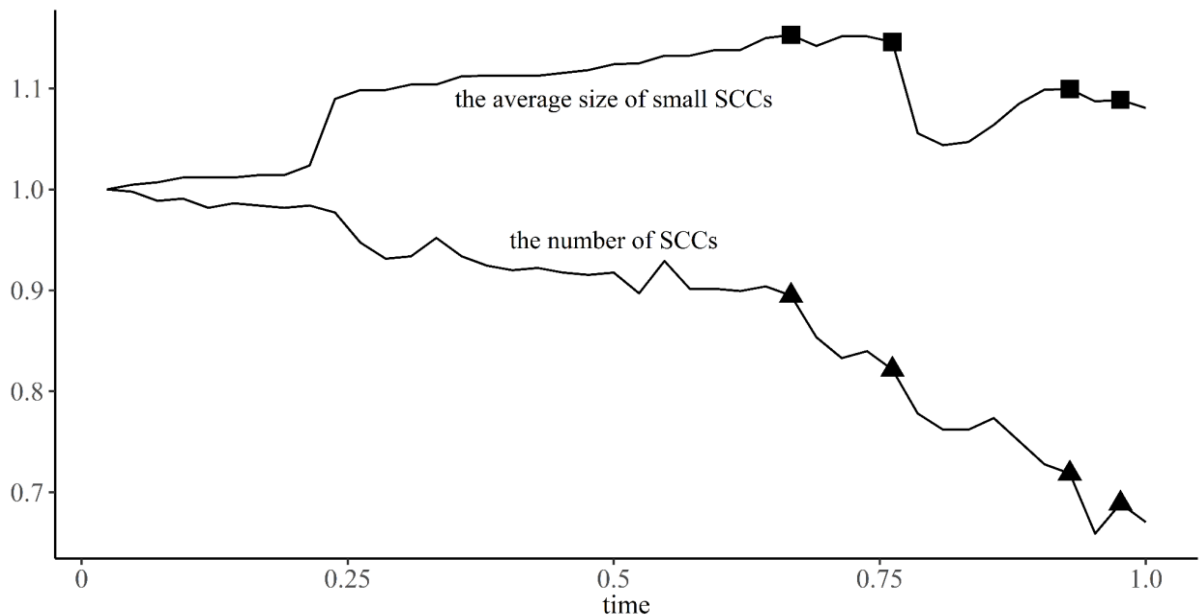

Fig. 6. Changes in the average size of small SCCs and the number of SCCs. The values are normalized with respect of each initial value. Following Newman [51], an isolated occupation is treated as an SCC of size 1. Closed dots indicate the moments when the GSCC growth accelerates.

The acceleration of structural changes through the scenario needs to be interpreted in relation to both the number and the categories of occupations participating in the GSCC. In the early stage of the scenario, the fragmented structure in the occupation network changes only marginally. As a result, the limited number of similar occupations exclusively form their own small SCCs. This suggests that occupational mobility may be constrained to the small number of similar occupations with premature automation. In the later stage, the structural changes become more significant with relatively little progression of the scenario. As fragmented SCCs integrate, the growth of the GSCC—reflected by an increasing number of occupations participating in it—accelerates. Simultaneously, occupations from all four categories begin to participate in the GSCC. These changes imply that as automation progresses, the expansion of the boundaries of smooth occupational mobility accelerates, enabling transitions between heterogeneous occupations.

*4.3 Mechanisms behind Structural Changes under Skill Automation Scenario*

The acceleration of structural changes in the occupation network follows Chang et al. [27]. However, the skill automation scenario in this study differs in two respects. First, the network in Chang et al. [27] is an undirected unipartite network, whereas the occupation network is a directed network that is the projection of the occupation–skill bipartite network onto the occupation mode. Second, and more importantly, they emphasize the role of added bridges integrating components, whereas the role of skill automation in integrating the SCCs in the occupation network remains unclear. In other words, the process of integrating SCCs and accelerated structural changes in the skill automation scenario is complex. We now explore the role of skill automation in integrating the SCCs, which triggers the acceleration of structural changes in the occupation network.

According to the discussion in Subsection 4.1, the specific skills determine the fragmented structure of the occupation network. These skills can be considered as bottlenecks to mobility across SCCs [19]. Therefore, the structural changes in an occupation network depend on whether automation substitutes specific skills or not. Fig. 7, which shows that skills can be broadly grouped into three categories based on $automatability_s$ and $specificity_s$, mirrors the findings from previous literature. Automatable skills are generally related to perception and manipulation, reflecting the findings of Frey and Osborne [5]. Specific skills are physical, sensory, and psychomotor abilities, and knowledge, whereas non-specific skills are generally basic and social skills, and cognitive abilities. This aligns with the findings of Alabdulkareem et al. [35] and Hosseinioun et al. [37]. Table C1 in Appendix C shows the measured value of $automatability_s$ and $specificity_s$.

In detail, among specific skills, those in the first quadrant—"Production and Processing," "Auditory Attention," "Repairing," "Dynamic Strength," and "Food Production"—have positive automatability values, whereas those in the second quadrant—"Fine Arts," "Medicine and Dentistry," "Science," "Engineering and Technology," and "Law and Government"—have negative automatability values. The non-specific and non-automatable skills in the third quadrant include "Complex Problem Solving," "Writing," "Education and Training," "Instructing," and "Negotiation." Amplifying prior literature [4,5,35,37], this result implies that automation substitutes the subset of specific skills, which is the bottleneck to mobility across SCCs.

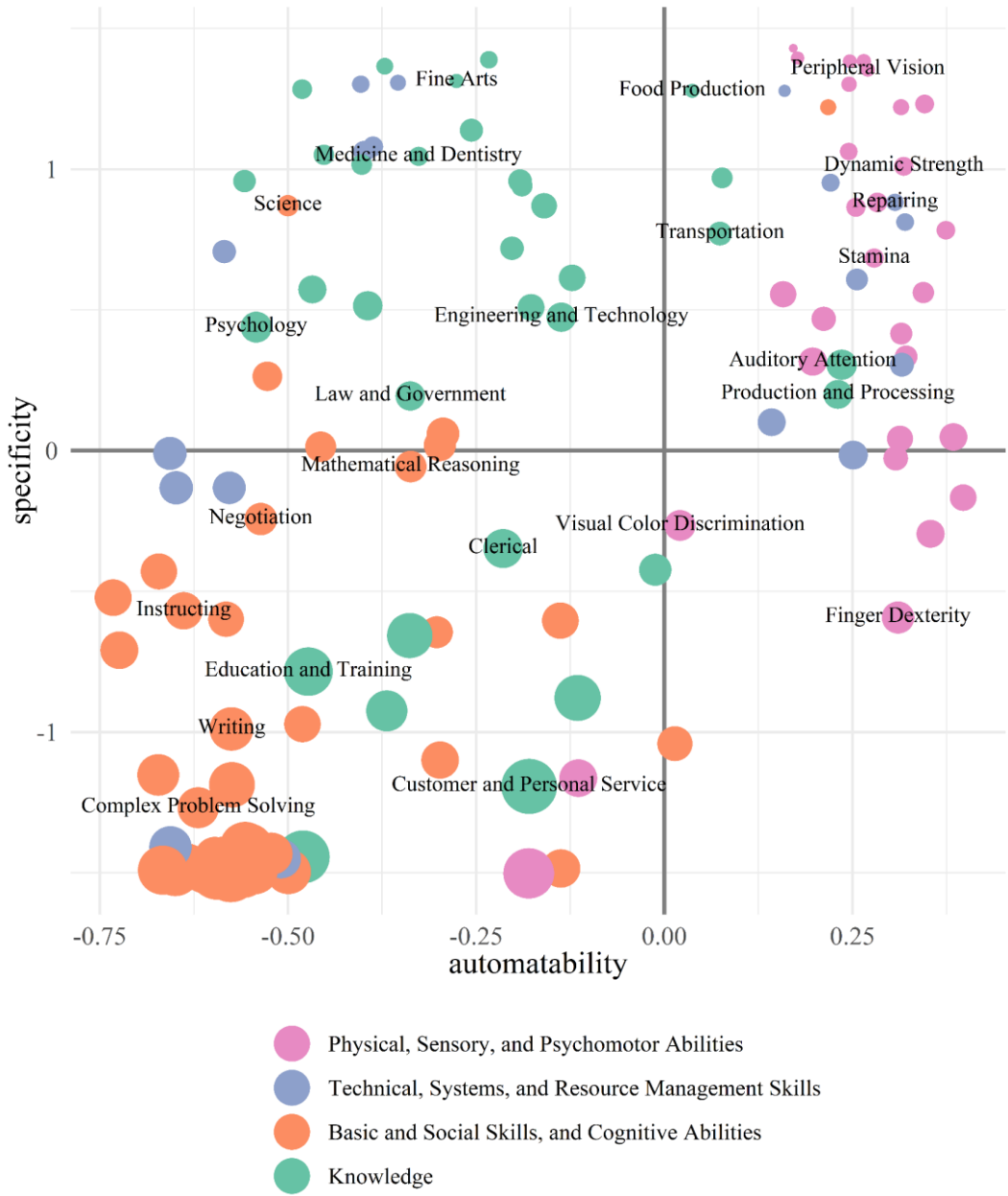

Fig. 7. The relationship between automatability and specificity of skills. The size of the points represents the skill generality as measured by Hosseinioun et al. [37]. The size of these points increases as these are located further down the y-axis due to their smaller specificity values. The specificity in this study and Hosseinioun et al.'s generality of skills is negatively correlated ($\rho = -0.95$), validating that our measure of specificity is consistent with prior works (see Fig. C1).

To understand the mechanism by which skill automation triggers the acceleration of structural changes in the occupation network, the directionality of links in this network should be considered. An SCC has cyclic paths, meaning that SCCs cannot integrate solely through out-directed one-way links. For them to integrate and form a new one, immediate or mediated two-way links are required. A one-way link between two SCCs indicates an asymmetric relationship, implying that the mobility that contradicts the directionality of the one-way link has a bottleneck. If this one-way asymmetric relationship evolves into a two-way symmetric relationship between SCCs, whether immediate or mediated, smooth mobility across the SCCs becomes feasible.

Fig. 8 shows an example where the removal of a specific skill triggers the integration of SCCs. In left panel, numerous one-way links are established from the GSCC to small SCC, implying the bottleneck of mobility from the small SCC to the GSCC. In right panel, when "Production and Processing" skill is removed from the skillset of "Management Analysts," a single link from "Market Research Analysts and Marketing Specialists" to "Management Analysts" is generated, thereby evolving into a two-way link between the two occupations. Subsequently, the boundaries of smooth mobility between "Market Research Analysts and Marketing Specialists," "Training and Development Specialists," and "Personal Financial Advisors" expand to occupations in the GSCC. This change shows that the removal of only one skill triggers the integration of two SCCs and the emergence of a new one. The integration of SCCs following this mechanism occurs repeatedly in the skill automation scenario. As these changes accumulate, they trigger the acceleration of significant structural changes in the occupation network.

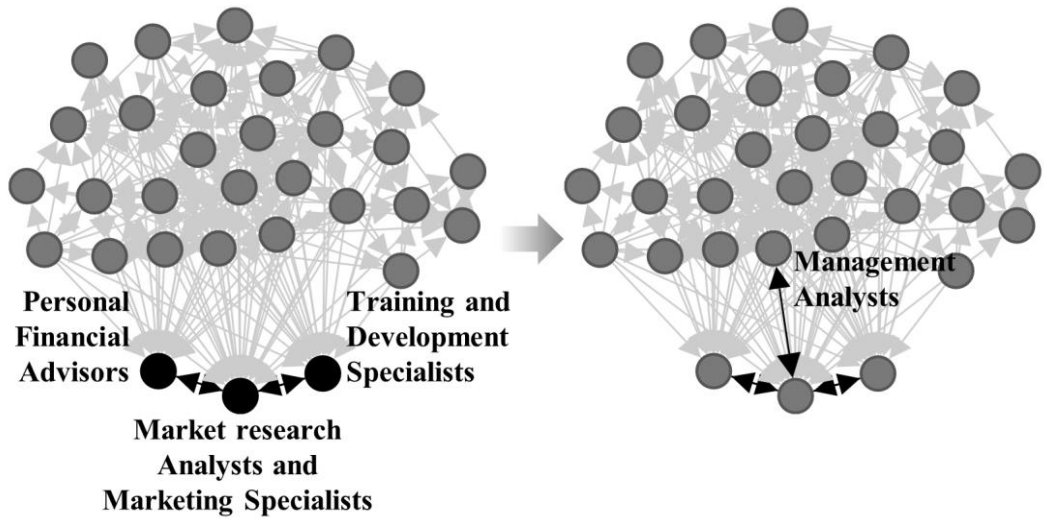

Fig. 8. The integration of GSCC and small SCC in the occupation network.

The mechanism by which skill automation triggers structural changes in the occupation network can be summarized as follows. Automation substitutes for the subset of specific skills that determine the fragmented structure in the network. As several specific skills are removed, the one-way links, which represent bottlenecks to occupational mobility across the SCCs, evolve into two-way links that integrate SCCs. As these changes accumulate, skill automation eventually accelerates structural changes in the occupation network.

## 5. Discussion

The debate surrounding technological unemployment has expanded to network studies proposing occupational mobility as a coping strategy for workers facing automation. However, previous studies have two prominent limitations. First, the cause of smooth mobility across dissimilar occupations remains implicit [18], with a focus on projected occupation networks. Second, despite the seminal literature [5,6] considering the impact of automation on skills, there is no focus on automatabilities of skills. The likelihood that the alteration of skillsets in occupations due to technological changes [4,7,22,24] would lead to the restructuring of the occupation network has often been overlooked.

This study addresses these limitations by proposing the skill automation view and simulating its impact on the occupation network, which is constructed by using O*NET data, through the sequential removal of automatable skills. To highlight the skills, the automatabilities of occupations estimated by Frey and Osborne [5] are translated to automatabilities of skills. Moreover, this study investigates not only the projected occupation network but also the occupation–skill bipartite network, which indicates the bundle of skills in each occupation, to directly reveal the underlying bottlenecks to smooth mobility.

Our findings are threefold. First, specific skills exclusively shared between similar occupations are explicitly identified as the cause of the fragmented structure in the occupation network, which remains implicit in the literature. Examining the occupation network, which projects asymmetric relationships between skillsets onto the occupation mode, can only capture the structure, not its underlying cause. In this study, the heterogeneous distribution of specific skills across a limited number of occupations is presented by concurrently observing the skillsets in occupations through the occupation–skill network. This approach directly reveals that certain skills determine the fragmented structure of the occupation network.

Second, the structure of the occupation network changes through the skill automation scenario. This change can be explained by percolation theory [27–29]. The structural changes are initially insignificant while the small SCCs grow, but they eventually accelerate as the SCCs integrate. This indicates that multiple small SCCs grow and then integrate with the GSCC, contributing to the acceleration of structural changes in the network. Consequently, a large number of heterogeneous occupations participate in the GSCC. These changes in the scenario suggest that as automation progresses, the boundaries of smooth mobility will expand to include a relatively large number of heterogeneous occupations.

Third, the complex mechanisms behind the structural changes in the scenario illustrated. The relationship between automatability and specificity of skills reveals that automation substitutes for the subset of specific skills, which function as bottlenecks to mobility across dissimilar occupations, particularly those related to perception and manipulation [5]. As these are mitigated, the relationships between SCCs evolve, triggering the acceleration of structural changes in the occupation network. These mechanisms show that skill automation will create shortcuts toward diverse occupations, allowing smooth mobility between heterogeneous occupations.

The study makes theoretical as well as practical contributions by showing that skill automation drives structural changes in the occupation network. The theoretical contributions are as follows. Our proposed skill automation view suggests that automation substitutes for skills, not occupations. The intangibility of skills presents a challenge for analysis, whereas occupations are more tangible because labor data is generally collected based on occupational classifications [4,24,54]. However, focusing on occupations is an approach to understanding labor dynamics, but not a dogma. In this respect, it is more appropriate to translate the automatabilities estimated by Frey and Osborne [5], which are matched to occupation codes, into the skill level. This enables the second theoretical contribution of this study. The skill automation scenario is an attempt to respond to the research agenda suggested by Frank et al. [4] on changes in labor trends due to technological change. The simulation approach that explains structural changes in the network according to percolation theory is an alternative way for understanding new labor trends, such as the employment growth driven by automation [38–40,55], contrary to conventional wisdom. In this sense, the skill automation scenario presented in this paper makes theoretical contributions by proposing a framework for understanding the complex mechanisms of emerging labor trends, which are too intricate to elucidate through a static occupation network.

Our practical contributions are as follows. This study implies that workers facing the risk of technological unemployment can find new uses for their labor in a broader range with continuous automation. Whereas this implication may seem paradoxical, it is straightforward to understand with an emphasis on occupation as a bundle of skills [4,6–8,20–24] and that automation substitutes for the subset of specific skills, not occupations. If the demand of specific skills, which are exclusively required by similar occupations, decreases due to automation, the bottlenecks to occupational mobility will be mitigated. Consequently, the burden required for workers to find new uses for their labor could significantly decrease, especially with the formation of shortcuts to occupational mobility. Nevertheless, we tend to focus solely on the loss of own specific skills. Drawing on Mokyr et al. [10], although the impact of technological changes on labor may be disruptive in the short term, it can eventually lead to long term beneficial outcomes. If individual workers understand automation as a process of delegating difficult and burdensome works previously performed by them to machines, automation will be seen not simply as a threat of job loss but also as an opportunity to find new uses for their labor across a wider range. Moreover, policymakers need to implement appropriate reskilling or labor redeployment policies with a long-term perspective to assist individual workers in coping with technological changes [11–15,24].

## 6. Conclusions

This study proposes the skill automation view for understanding labor trends by highlighting skills rather than focusing solely on occupations. To address the limitations of existing studies, which primarily consider static networks and the automatability of occupations, this study simulates structural changes in the occupation network through a skill automation scenario by sequentially removing skills from the network. The findings from this scenario, which show the integration of fragmented components in the occupation network, challenge the conventional wisdom of automation as a threat to job loss. Therefore, this study contributes to the theoretical debate on technological unemployment by proposing an alternative approach.

The framework we present offers a practical way to swiftly find, across a wider range, new uses for individual workers' labor in response to technological change. The discussions also inform policy to mitigate technological unemployment by understanding how the automation of specific skills can eventually create shortcuts to facilitate labor redeployment in the long term.

This study has two primary limitations that offer avenues for future research. First, a single set of automatability estimates was used. Although Frey and Osborne [5] estimated the automatability of occupations by considering a comprehensive range of machines, including robots and artificial intelligence, their methods incorporate both objective and subjective factors. To enhance the robustness of the approach in this study, future studies can explore simulations using alternative automatability estimates. These include those focused particularly on the impact of artificial intelligence on labor [7]. Second, this study highlights specific skills, but not general skills, even though both are of interest to human capital scholars. Specific and general skills have trade-off in relation to higher return and mobility [26]. A growing body of literature utilizing network analysis has discussed the contribution of specific and general skills to labor dynamics [37]. Future work, based on network science, may better describe how workers can utilize general skills to cope with the ongoing impact of technological change on the labor market. Third, the productivity effect of technology on labor was excluded in this study while citing the estimates of Frey and Osborne [5], which considered the displacement effect. Technology not only substitutes for labor but also complements it [9]. If future research incorporates estimates that reflect the productivity effect of technology in our approach, such as the increasing demand for new skills or jobs [56], a more dynamic model could be suggested.

**Declaration of interest**

The authors have no competing interests to declare that are relevant to the content of this article.

**Funding**

This work was supported by the National Research Foundation of Korea (NRF) grant funded by the Korean government (MSIT)) (grant numbers NRF-2022R1A2C1091917, NRF-2023S1A5B5A19093480).

**Author contributions**

**Soohyoung Lee:** Conceptualization; Data curation; Formal analysis; Investigation; Methodology; Resources; Software; Validation; Visualization; Writing - original draft; Writing - review & editing

**Dawoon Jeong:** Conceptualization; Formal analysis; Investigation; Methodology; Supervision; Validation; Visualization; Writing - original draft; Writing - review & editing

**Jeong-Dong Lee:** Conceptualization; Funding acquisition; Project administration; Resources; Supervision; Writing - review & editing

**References**

1. J.M. Keynes, Economic possibilities for our grandchildren. In: J.M. Keynes (Ed.), Essays in Persuasion. Palgrave MacMillan, 1931, pp. 321–32.

2. J.D. Dworkin, 2019, Network-driven differences in mobility and optimal transitions among automatable jobs. R. Soc. Open Sci. 6, 182124. https://doi.org/10.1098/rsos.182124.

3. A. Christenko, 2022, Automation and occupational mobility: a task and knowledge-based approach. Technol. Soc. 70, 101976. https://doi.org/10.1016/j.techsoc.2022.101976.

4. M.R. Frank, D. Autor, J.E. Bessen, E. Brynjolfsson, M. Cebrian, D.J. Deming, et al., 2019, Toward understanding the impact of artificial intelligence on labor. Proc. Natl. Acad. Sci. USA 116(14), 6531–39. https://doi.org/10.1073/pnas.1900949116.

5. C.B. Frey, M.A. Osborne, 2017, The future of employment: how susceptible are jobs to computerisation? Technol. Forecast. Soc. Change.114, 254–80. https://doi.org/10.1016/j.techfore.2016.08.019.

6. D.H. Autor, F. Levy, R.J. Murnane, 2003, The skill content of recent technological change: An empirical exploration. Q. J. Econ. 118(4), 1279–333. https://doi.org/10.1162/003355303322552801.

7. E. Brynjolfsson, T. Mitchell, D. Rock, 2018, What can machines learn and what does it mean for occupations and the economy? AEA Pap. Proc. 108, 43–7. https://doi.org/10.1257/pandp.20181019.

8. L. Nedelkoska, D. Diodato, F. Neffke, 2018, Is our human capital in general enough to withstand the current wave of technological change? CID Working Papers 93a, Center for International Development at Harvard University.

9. D. H. Autor, 2015, Why are there still so many jobs? The history and future of workplace automation. J. Econ. Perspect. 29(3), 3–30. https://doi.org/10.1257/jep.29.3.3.

10. J. Mokyr, C. Vickers, N.L. Ziebarth, 2015, The history of technological anxiety and the future of economic growth: is this time different? J. Econ. Perspect. 29(3), 31–50. https://doi.org/10.1257/jep.29.3.31.

11. Y. Yeo, W.S. Hwang, J.D. Lee, 2023, The shrinking middle: exploring the nexus between information and communication technology, growth, and inequality. Technol. Econ. Dev. Economy. 29(3), 874–901. https://doi.org/10.3846/tede.2023.18713.

12. K. Hötte, M. Somers, A. Theodorakopoulos, 2023, Technology and jobs: a systematic literature review. Technol. Forecast. Soc. Change. 194, 122750. https://doi.org/10.1016/j.techfore.2023.122750.

13. S. McGuinness, K.P. Pouliakas, P. Redmond, 2023, Skills-displacing technological change and its impact on jobs: challenging technological alarmism? Econ. Innov. New Technol. 32(3), 370–92. https://doi.org/10.1080/10438599.2021.1919517.

14. Y. Yeo, J.D. Lee, 2020, Revitalizing the race between technology and education: investigating the growth strategy for the knowledge-based economy based on a CGE analysis. Technol. Soc. 62, 101295. https://doi.org/10.1016/j.techsoc.2020.101295

15. Y. Yeo, J.D. Lee, S. Jung, 2023, Winners and losers in a knowledge-based economy: investigating the policy packages for an inclusive growth based on a computable general equilibrium analysis of Korea. J. Asia Pac. Econ. 28(2), 420–56. https://doi.org/10.1080/13547860.2021.1982193.

16. F. Neffke, L. Nedelkoska, S. Wiederhold, 2024, Skill mismatch and the costs of job displacement. Res. Policy. 53(2), 104933. https://doi.org/10.1016/j.respol.2023.104933.

17. E. Moro, M.R. Frank, A. Pentland, A. Rutherford, M. Cebrian, I. Rahwan, 2021, Universal resilience patterns in labor markets. Natur. Commun. 12, 1972. https://doi.org/10.1038/s41467-021-22086-3.

18. F.M.H. Neffke, A. Otto, A. Weyh, 2017, Inter-industry labor flows. J. Econ. Behav. Organ. 142, 275–92. https://doi.org/10.1016/j.jebo.2017.07.003

19. J. Hervé, 2023, Specialists or generalists? Cross-industry mobility and wages. Labour Econ. 84, 102391. https://doi.org/10.1016/j.labeco.2023.102391.

20. L. Nedelkoska, F. Neffke, 2019, Skill mismatch and skill transferability: Review of concepts and measurements. Pap. Evol. Econ. Geogr. 19(21). The Growth Lab. Harvard University.

21. S. Sampson, 2021, A strategic framework for task automation in professional services. J. Serv. Res. 24(1), 122–40. https://doi.org/10.1177/1094670520940407.

22. F. Stephany, O. Teutloff, 2024, What is the price of a skill? The value of complementarity. Res. Policy. 53(1), 104898. https://doi.org/10.1016/j.respol.2023.104898.

23. C. Robinson, 2018, Occupational mobility, occupation distance, and specific human capital. J. Hum. Resour. 53(2), 513–551. https://doi.org/10.3368/jhr.53.2.0814-6556R2.

24. M. Duan, Y. Hou, B. Zhang, C. Chen, Y. Sun, Y. Luo, T. Tan, 2024, Skill sets and wage premium: a network analysis based on Chinese agriculture online job offers. Technol. Forecast. Soc. Change. 201, 123260. https://doi.org/10.1016/j.techfore.2024.123260.

25. E.P. Lazear, 2009, Firm-specific human capital: a skill-weights approach. J. Polit. Econ. 117(5), 914–40. https://doi.org/10.1086/648671.

26. C. Eggenberger, M. Rinawi, U. Backes-Gellner, 2018, Occupational specificity: a new measurement based on training curricula and its effect on labor market outcomes. Labour Econ. 51(April), 97–107. https://doi.org/10.1016/j.labeco.2017.11.010.

27. S. Chang, J. Lee, J. Song, 2023, Giant cluster formation and integrating role of bridges in social diffusion. Strateg. Manag. J. 44(12), 2950–85. https://doi.org/10.1002/smj.3527.

28. D. Stauffer, A. Aharony, Introduction to Percolation Theory, Second ed., Taylor & Francis, 1991. https://doi.org/10.1201/9781315274386.

29. H.O. Peitgen, H. Jürgens, D. Saupe, Chaos and Fractals: New Frontiers of Science, Second ed., Springer, 2004.

30. R. Albert, H. Jeong, A.L. Barabasi, 2000, Error and attack tolerance of complex networks. Nature. 406(6794), 378–82. https://doi.org/10.1038/35019019.

31. W. Jo, D. Chang, M. You, G.H. Ghim, 2021, A social network analysis of the spread of covid-19 in South Korea and policy implications. Sci. Rep. 11(1), 8581. https://doi.org/10.1038/s41598-021-87837-0.

32. F. Neffke, M. Henning, 2013, Skill relatedness and firm diversification. Strateg. Manag. J. 34(3), 297–316. https://doi.org/10.1002/smj.2014.

33. M. Poletaev, C. Robinson, 2008, Human capital specificity: evidence from the dictionary of occupational titles and displaced worker surveys, 1984–2000. J. Labor Econ. 26(3), 387–420. https://doi.org/10.1086/588180.

34. C. Gathmann, U. Schönberg, 2010, How general is human capital? A task-based approach. J. Labor Econ. 28(1), 1–49. https://doi.org/10.1086/649786.

35. A. Alabdulkareem, M.R. Frank, L. Sun, B. Alshebli, C. Hidalgo, I. Rahwan, 2018. Unpacking the polarization of workplace skills. Sci. Adv. 4(7), eaao6030. https://doi.org/10.1126/sciadv.aao6030.

36. R. Boschma, 2005, Proximity and innovation: a critical assessment. Reg. Stud. 39(1), 61–74. https://doi.org/10.1080/0034340052000320887.

37. M. Hosseinioun, F. Neffke, Z. Letian, H. Youn, 2024, Nested skills in labor ecosystems: a hidden dimension of human capital. Ithaca: Cornell University Library, arXiv.org.

38. D. Klenert, E. Fernández-Macías, J.I. Antón, 2023, Do robots really destroy jobs? Evidence from Europe. Econ. Ind. Democr. 44(1), 280–316.
https://doi.org/10.1177/0143831X211068891

39. Q. Zhang, F. Zhang, Q. Mai, 2023, Robot adoption and labor demand: A new interpretation from external competition. Technol. Soc. https://doi.org/10.1016/j.techsoc.2023.102310.

40. M. Xue, X. Cao, X. Feng, B. Gu, Y. Zhang, 2022, Is college education less necessary with AI? Evidence from firm-level labor structure changes. J. Manag. Inf. Syst. 39(3), 865–905. https://doi.org/10.1080/07421222.2022.2096542

41. M. Downey, 2021, Partial automation and the technology-enabled deskilling of routine jobs. Labour Econ. 69, 101973. https://doi.org/10.1016/j.labeco.2021.101973

42. C.A. Hidalgo, B. Klinger B., A.L. Barabási, R. Hausmann, 2007, The product space conditions the development of nations. Science. 317, 5837. https://doi.org/10.1126/science.1144581

43. Kogler D.F., D.L. Rigby, I. Tucker, 2013, Mapping knowledge space and technological relatedness in US cities. Eur. Plan. Stud. 21(9), 1374–91. https://doi.org/10.1080/09654313.2012.755832

44. Y. Chun, J. Hur, J. Hwang, 2024, AI technology specialization and national competitiveness. PLoS One. https://doi.org/10.1371/journal.pone.0301091

45. M.R. Guevara, D. Hartmann, M. Aristarán, M. Mendoza, C.A. Hidalgo, 2016, The research space: using career paths to predict the evolution of the research output of individuals, institutions, and nations. Scientometrics. 109, 1695–709. https://doi.org/10.1007/s11192-016-2125-9

46. R. Hausmann, B. Klinger, 2007, The structure of the product space and the evolution of comparative advantage. CID Working Paper No. 146, Harvard University. http://nrs.harvard.edu/urn-3:HUL.InstRepos:42482358

47. R. Boschma, 2017, Relatedness as driver of regional diversification: a research agenda. Reg. Stud. 51(3), 351–64. https://doi.org/10.1080/00343404.2016.1254767

48. J. Memmott, N.M. Waser, M.V. Price, 2004, Tolerance of pollination networks to species extinctions. Proc. R. Soc. Lond. B. 271, 2605–11. https://doi.org/10.1098/rspb.2004.2909

49. Evans D.M., Pocock M.J.O., J. Memmott, 2013, The robustness of a network of ecological networks to habitat loss. Ecol. Lett. 16(7), 844–52. https://doi.org/10.1111/ele.12117

50. J. Park, J. Kim, 2022, A data-driven exploration of the race between human labor and machines in the 21st century. Commun. ACM. 65(5), 79–87. https://doi.org/10.1145/3488376

51. M.E.J. Newman, Networks, Oxford University Press, 2018.

52. A. Elliott, A. Chiu, M. Bazzi, G. Reinert, M. Cucuringu, 2020, Core-periphery structure in directed networks. Proc. R. Soc. A. 476, 20190783. https://doi.org/10.1098/rspa.2019.0783

53. J.D. Lee, D. Jeong, E. Y. Jung, Y. Kim, J. Kim, Y. He, S. Choi, 2022. Mapping the evolutionary pattern of mobile products: a phylogenetic approach. IEEE Trans. Eng. Manag. 71, 4776–90, https://doi.org/10.1109/TEM.2022.3214489

54. D.J. Deming, K. Noray, 2020, Earnings dynamics, changing job skills, and STEM careers. Q. J. Econ. 135(3), 1965–2005. https://doi.org/10.1093/qje/qjaa021

55. G. Domini, M. Grazzi, D. Moschella, T. Treibich, 2021. Threats and opportunities in the digital era: automation spikes and employment dynamics. Res. Policy. 50(7), 104137. https://doi.org/10.1016/j.respol.2020.104137

56. T. Berger, C.B. Frey, 2016, Did the computer revolution shift the fortunes of U.S. cities? Technology shocks and the geography of new jobs. Reg. Sci. Urban Econ. 57, 38–45. https://doi.org/10.1016/j.regsciurbeco.2015.11.003

57. A. Clauset, C. Shalizi, M.E.J. Newman, 2009, Power-law distributions in empirical data. SIAM Rev. 51(4). https://doi.org/10.1137/070710111

58. A. Clauset, 2018. Trends and fluctuations in the severity of interstate wars. Sci. Adv. 4(2). https://doi.org/10.1126/sciadv.aao3580.

## Appendix A. Determining the threshold of $T_{o,o'}$ to build the occupation network

$T_{o,o'}$ from 505 to 504 occupations represents transition "potential." Alabdulkareem et al. [35] suggested that transitions between occupations with disparate skillsets are improbable, indicating low transition potential. Thus, constructing the occupation network solely with links surpassing a specific threshold of $T_{o,o'}$ enhances its realism in portraying real-world occupational mobility.

To establish this threshold, we construct a real-world occupational transition network using data on occupation-to-occupation transitions from the IPUMS-CPS dataset (January 2018–January 2023).[5] Matching Census Occupation Codes (COC) in IPUMS-CPS data to Standard Occupation Codes (SOC) in the O*NET data is the initial step. We preprocess the "2018 occupation code list and crosswalk" [6] from the U.S. Census Bureau to ensure accurate code matching. Whereas O*NET data utilizes 8-digit O*NET-SOC, the crosswalk list provides 6-digit SOC. We align the 6-digit SOC from O*NET with COC and average the skill importance values based on COC, yielding 505 Census Occupation Codes.

The network's structure is typically depicted by its degree distribution. Newman [51] suggested assessing the joint distribution of in and out degrees for an accurate representation of a directed network's structure. Hence, we employ the Kolmogorov–Smirnov test to compare the joint distributions of in and out degrees of the two networks. This test, utilized in the literature to discern network structures, calculates the maximum distance ($D$) between two cumulative distributions [31,57,58]. A smaller $D$ implies greater similarity between the distributions. We estimated $D$ by comparing the joint degree distribution of a real-world occupational transition network with those of occupation networks constructed using $T_{o,o'}$ above given $\tilde{\theta}$. Consequently, we determine the threshold $\theta$ for $T_{o,o'}$ as the value resulting in the smallest $D$. Fig. A1 indicates that the occupation network constructed with links above 0.94 yields the smallest $D$ value. Therefore, we establish 0.94 as the threshold for $T_{o,o'}$.

---

[5] See https://cps.ipums.org/cps/
[6] See https://www.census.gov/topics/employment/industry-occupation/guidance/code-lists.html

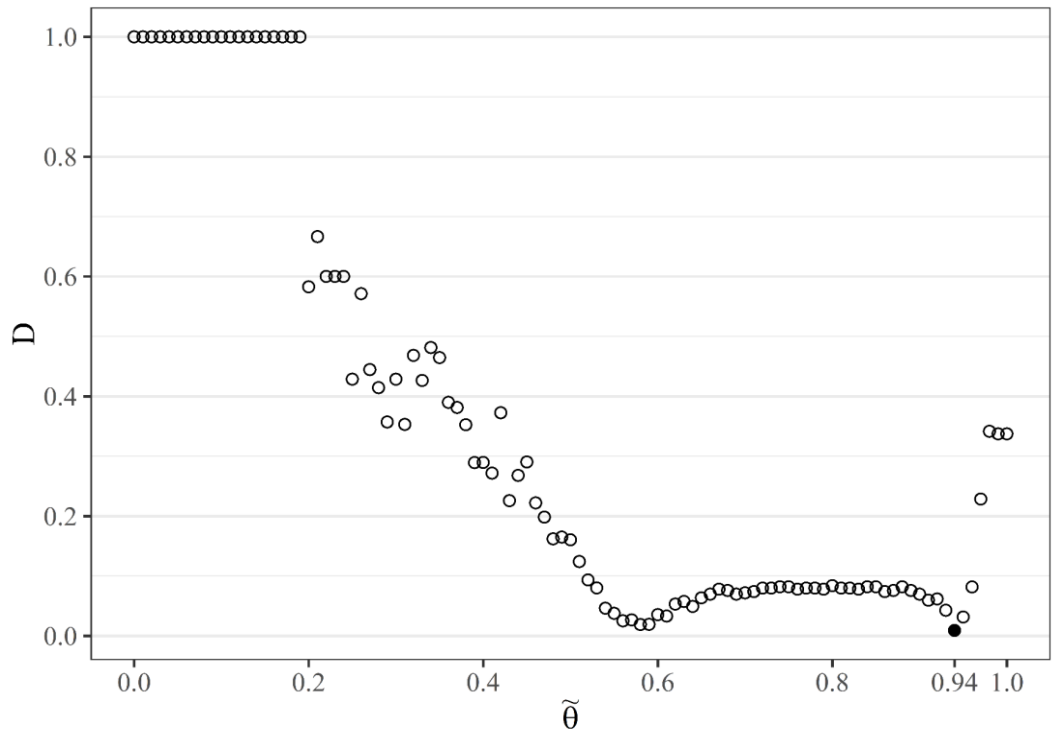

Fig. A1. KS statistic ($D$) obtained by comparing occupation networks with different $\tilde{\theta}$.

*Appendix B. The occupation–skill network*

Following Memmott, Waser, and Price [48], we visualize the occupation–skill network in a matrix form in Fig. B1. The axes are sorted by the number of links that occupations and skills have. The distribution of skills is heterogeneous across occupations [25,26]. This approach can more intuitively represent the specificity of skills.

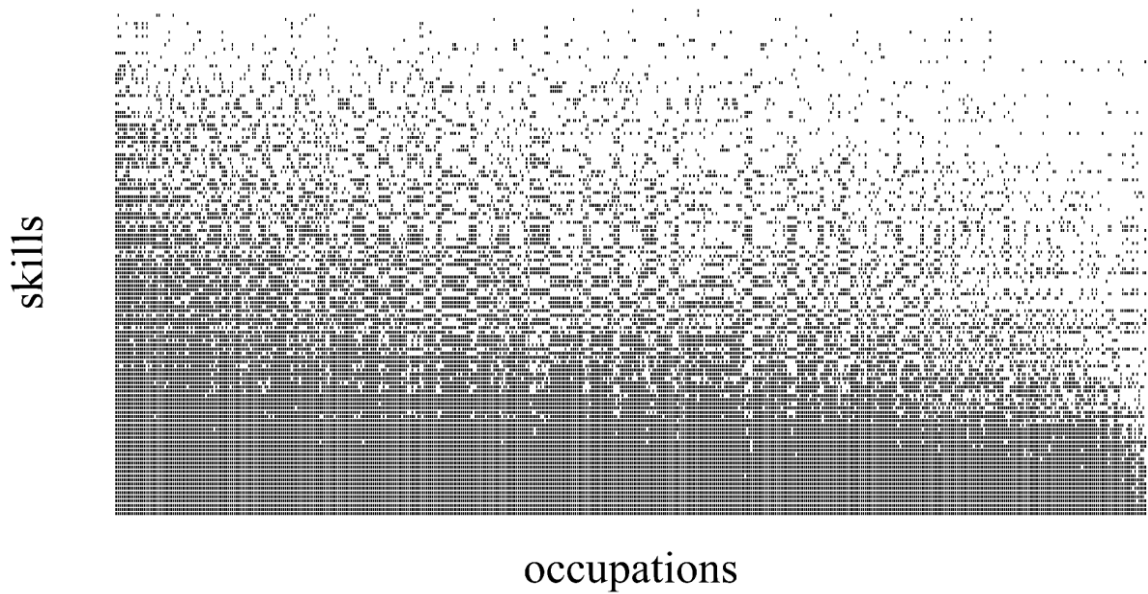

Fig. B1. The occupation-skill network as a matrix form. A black tile indicates the link between an occupation and a skill ($R_{o,s} = 1$).

## Appendix C. Automatability and specificity of skills

Table C1. Automatability and specificity of skills.

| Skill | $automatability_s$ | $specificity_s$ |
|---|---|---|
| Manual Dexterity | 0.4 | -0.17 |
| Control Precision | 0.38 | 0.05 |
| Rate Control | 0.37 | 0.78 |
| Arm-Hand Steadiness | 0.35 | -0.3 |
| Wrist-Finger Speed | 0.35 | 1.23 |
| Reaction Time | 0.34 | 0.56 |
| Static Strength | 0.32 | 0.33 |
| Equipment Maintenance | 0.32 | 0.81 |
| Dynamic Strength | 0.32 | 1.01 |
| Operation and Control | 0.32 | 0.3 |
| Extent Flexibility | 0.31 | 0.42 |
| Speed of Limb Movement | 0.31 | 1.22 |
| Trunk Strength | 0.31 | 0.04 |
| Finger Dexterity | 0.31 | -0.59 |
| Multi-limb Coordination | 0.31 | -0.03 |
| Repairing | 0.31 | 0.88 |
| Response Orientation | 0.28 | 0.88 |
| Stamina | 0.28 | 0.68 |
| Peripheral Vision | 0.27 | 1.35 |
| Sound Localization | 0.26 | 1.38 |
| Troubleshooting | 0.26 | 0.61 |
| Gross Body Coordination | 0.25 | 0.86 |
| Operation Monitoring | 0.25 | -0.02 |
| Night Vision | 0.25 | 1.38 |
| Glare Sensitivity | 0.25 | 1.3 |
| Gross Body Equilibrium | 0.24 | 1.06 |
| Mechanical | 0.24 | 0.3 |
| Production and Processing | 0.23 | 0.2 |
| Equipment Selection | 0.22 | 0.95 |
| Spatial Orientation | 0.22 | 1.22 |
| Depth Perception | 0.21 | 0.47 |
| Auditory Attention | 0.2 | 0.32 |
| Explosive Strength | 0.18 | 1.4 |
| Dynamic Flexibility | 0.17 | 1.43 |
| Installation | 0.16 | 1.28 |
| Hearing Sensitivity | 0.16 | 0.56 |
| Quality Control Analysis | 0.14 | 0.1 |
| Building and Construction | 0.08 | 0.97 |
| Transportation | 0.07 | 0.77 |
| Food Production | 0.04 | 1.28 |
| Visual Color Discrimination | 0.02 | -0.27 |
| Perceptual Speed | 0.01 | -1.04 |
| Public Safety and Security | -0.01 | -0.42 |
| Far Vision | -0.11 | -1.16 |
| Mathematics (knowledge) | -0.12 | -0.88 |
| Design | -0.12 | 0.61 |
| Engineering and Technology | -0.14 | 0.47 |
| Selective Attention | -0.14 | -1.48 |
| Visualization | -0.14 | -0.6 |
| Chemistry | -0.16 | 0.87 |
| Sales and Marketing | -0.18 | 0.51 |
| Customer and Personal Service | -0.18 | -1.19 |
| Near Vision | -0.18 | -1.5 |
| Telecommunications | -0.19 | 0.94 |
| Physics | -0.19 | 0.96 |
| Economics and Accounting | -0.2 | 0.72 |
| Clerical | -0.21 | -0.35 |
| Foreign Language | -0.23 | 1.39 |
| Geography | -0.26 | 1.14 |
| Fine Arts | -0.28 | 1.31 |
| Mathematics_ | -0.29 | 0.06 |
| Flexibility of Closure | -0.3 | -1.1 |
| Number Facility | -0.3 | 0.02 |
| Time Sharing | -0.3 | -0.65 |
| Medicine and Dentistry | -0.33 | 1.05 |
| Mathematical Reasoning | -0.34 | -0.06 |
| Law and Government | -0.34 | 0.19 |
| Computers and Electronics | -0.34 | -0.66 |
| Programming | -0.35 | 1.31 |
| Administration and Management | -0.37 | -0.93 |
| History and Archeology | -0.37 | 1.37 |
| Management of Financial Resources | -0.39 | 1.08 |
| Personnel and Human Resources | -0.39 | 0.51 |
| Management of Material Resources | -0.4 | 1.06 |
| Biology | -0.4 | 1.02 |
| Technology Design | -0.4 | 1.3 |
| Therapy and Counseling | -0.45 | 1.05 |
| Speed of Closure | -0.46 | 0.01 |
| Communications and Media | -0.47 | 0.57 |
| Education and Training | -0.47 | -0.79 |
| English Language | -0.48 | -1.44 |
| Service Orientation | -0.48 | -0.97 |
| Philosophy and Theology | -0.48 | 1.28 |
| Information Ordering | -0.5 | -1.5 |
| Science | -0.5 | 0.87 |
| Time Management | -0.51 | -1.45 |
| Category Flexibility | -0.52 | -1.43 |
| Memorization | -0.53 | 0.26 |
| Negotiation | -0.54 | -0.24 |
| Psychology | -0.54 | 0.44 |
| Speech Recognition | -0.54 | -1.5 |
| Monitoring | -0.54 | -1.5 |
| Written Comprehension | -0.55 | -1.41 |
| Reading Comprehension | -0.56 | -1.41 |
| Sociology and Anthropology | -0.56 | 0.96 |
| Problem Sensitivity | -0.56 | -1.5 |
| Speech Clarity | -0.57 | -1.49 |
| Written Expression | -0.57 | -1.19 |
| Writing | -0.57 | -0.99 |
| Oral Comprehension | -0.58 | -1.5 |
| Management of Personnel Resources | -0.58 | -0.13 |
| Persuasion | -0.58 | -0.6 |
| Coordination | -0.58 | -1.44 |
| Operations Analysis | -0.58 | 0.71 |
| Oral Expression | -0.59 | -1.5 |
| Social Perceptiveness | -0.6 | -1.44 |
| Active Listening | -0.61 | -1.49 |
| Speaking | -0.61 | -1.49 |
| Complex Problem Solving | -0.62 | -1.27 |
| Inductive Reasoning | -0.64 | -1.48 |
| Instructing | -0.64 | -0.57 |
| Systems Analysis | -0.65 | -0.13 |
| Critical Thinking | -0.65 | -1.5 |
| Judgment and Decision Making | -0.66 | -1.41 |
| Systems Evaluation | -0.66 | -0.01 |
| Deductive Reasoning | -0.67 | -1.49 |
| Learning Strategies | -0.67 | -0.43 |
| Active Learning | -0.67 | -1.15 |
| Fluency of Ideas | -0.72 | -0.71 |
| Originality | -0.73 | -0.52 |

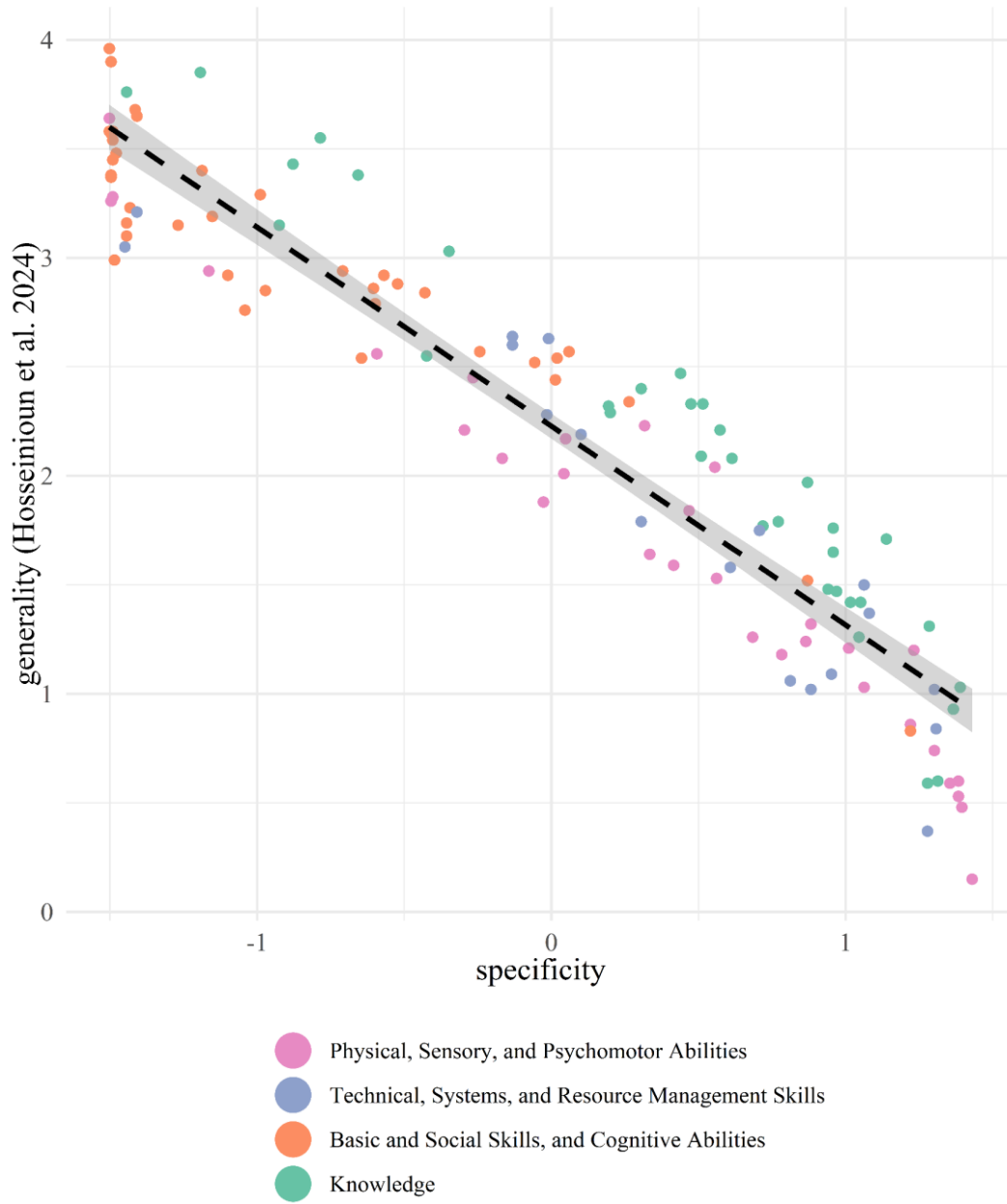

Fig. C1. The relationship between specificity and Hosseinioun et al.'s [37] generality of skills.